%Paper: gr-qc/9403046
%From: DIOSI@rmk530.rmki.kfki.hu
%Date: Wed, 23 Mar 1994 12:20 GMT+1

\magnification 1200
\def\kb{{\bf k}}\def\pb{{\bf p}}\def\qb{{\bf q}}
\def\pbi{{\bf p}_i}\def\pbf{{\bf p}_f}
\def\kbi{{\bf k}_i}\def\kbf{{\bf k}_f}\def\kbfi{{\bf k}_{fi}}
\def\kbia{{\bf k}_i^\ast}\def\kbfa{{\bf k}_f^\ast}\def\kbfia{{\bf k}_{fi}^\ast}
\def\kbiap{{\bf k}_i^{\ast'}}\def\kbfap{{\bf k}_f^{\ast'}}
\def\qbo{{\bf q}}\def\pbo{{\bf p}}\def\kboi{{\bf k}_i}
\def\E{{\cal E}}\def\P{{\cal P}}
\def\r{\rho}\def\re{\rho^{\E}}

{\it bulletin board ref.: gr-qc/9403046\hfill}
\bigskip
\centerline{\bf Quantum Master Equation of Particle in Gas Environment}
\bigskip
\centerline{Lajos Di\'osi$^{(a)}$}
\bigskip
\centerline{\it KFKI Research Institute for Particle and Nuclear Physics}
\centerline{H-1525 Budapest 114, POB 49, Hungary}
\bigskip
\centerline{Abstract}
The evolution of the reduced density operator $\r$
of Brownian particle
is discussed in single collision approach valid typically in low density gas
environments. This is the first succesful derivation of quantum friction
caused by {\it local} environmental interactions.
We derive a Lindblad master equation for $\r$,
whose generators are calculated from differential cross section
of a single collision between Brownian and gas particles, respectively.
The existence of thermal equilibrium for $\r$ is proved.
Master equations proposed earlier are shown to be particular cases of
our one.
\bigskip

Brownian motion is one of the most studied classical problem. It has drawn
special attention in the quantum case, too.
The irreversible dynamics of quantum particle interacting with a
stationary environment (e.g. a
thermalized reservoir or a stream of molecules, photons, phonons,
e.t.c.) is related
with many interesting fields in physics.
Nonetheless, the quantum counterpart of the classical
Brownian motion has not been casted into standard equations since
the quantum case is richer than the classical one, especially quantum
friction is not straightforward to include correctly.
In this Letter we derive quantum frictional equations arising from
{\it local} environmental interactions  typical for a gas environment.

In the classical fenomenological theory,
the momentum distribution $\r(\pb)$ of a
Brownian particle of mass $M$ satisfies the Fokker-Planck equation:
$$
{d\r(\pb)\over dt}=D_{pp}{\bf\nabla}^2\r(\pb)
                     +{\eta\over M}{\bf\nabla}(\pb\r(\pb))
\eqno(1)$$
where $D_{pp}$ is the coefficient of momentum diffusion and $\eta$ is the
friction constant. In particular, $D_{pp}=k_BT\eta$ if the environment is
in equilibrium at temperature $T$.
A naive quantum counterpart would be the following master equation [1]:
$$
{d\r\over dt}= -D_{pp}[\qbo,[\qbo,\r]]
                 -i{\eta\over2M}[\qbo,\{\pbo,\r\}]
\eqno(2)$$
where $\qb,\pb$ denote the position and, resp., the momentum {\it operators}
in interaction picture. This equation has been used
recently to describe the paradigm of environmental decoherence [2].
Eq.~(2), however, does not belong to the Lindblad class [3] and, consequently,
may violate the positivity of the
{\it density operator} $\r$ [4]. A detailed dynamic analysis of the quantum
Brownian motion in gas environment is thus needed to derive a
suitable quantum evolution equation of Lindblad class.

Let us first consider a simple (frictionless) classical kinetic model.
The momentum distribution of a Brownian particle,
interacting with the particles of the environment, satisfies
the following kinetic equation:
$$
{d\r(\pb)\over dt}=
        n_0\int dE d\Omega_i d\Omega_f k^2
        {d\sigma(\theta,E)\over d\Omega_f} \re(\kbi)
        \bigl(\r(\pb-\kbfi)-\r(\pb)\bigr)
\eqno(3)$$
where $n_0$ is the density of environmental particles,
$\re$ is their momentum distribution;
$d\sigma$ denotes the differential cross section of their scattering
on the Brownian particle, while $\kbfi=\kbf-\kbi$ is the difference
between the final and initial momenta of the scattered particle.
Further symbols are used in a standard way and need no special
explanation.
The Eq.~(3) is valid provided: the typical scattering time is
much smaller then the average period between subsequent collisions, the
mass $M$ of the Brownian particle is much bigger than the mass $m$ of the
environment particles, their interaction is spin independent, and
the influence of the Brownian particle on the environment's momentum
distribution $\re$ can be ignored.

Let us illustrate the additional issues
arising in the quantum case. Looking for a master equation for the
density operator $\r$, observe
the classical
Eq.~(3) would govern the evolution of its diagonal elements
according to the subsitution $\r(\pb)=<\pb|\r|\pb>$. For the
off-diagonal part of $\r$, however, we have no evolution rule.
Such quantum master equations, constructed after Pauli's pioneering one [5],
are nothing to do with quantum states which are not translation invariant.
A complete quantum kinetic equation must specify rules for $\r$'s
off-diagonal part as well.
Such an extension of the "Pauli" evolution Eq.~(3) would be
$$
{d\r\over dt}=
        n_0\int dE d\Omega_i d\Omega_f k^2
        {d\sigma(\theta,E)\over d\Omega_f} \re(\kbi)
        \left(e^{-i\kbfi\qbo}\r e^{i\kbfi\qbo}-\r\right).
\eqno(4)$$
This equation was, in fact, derived from the unitary
dynamics of the Brownian particle interacting with the gas environment
[6,7]).

The quantum master Eq.~(4) is not completely realistic since it
ignores dissipation (friction).
In this Letter we present a systematic derivation
of the dissipative quantum master equation. We start from
the unitary dynamics of the density operator $\r^{\P+\E}$ corresponding
to the Brownian particle $(\P)$ plus the environment $(\E)$
coupled to each other.
In interaction picture, a single collision between the Brownian and an
environmental particle transforms the initial state $\r_i^{\P+\E}$
into the final one $\r_f^{\P+\E}$ via the unitary scattering operator $S$:
$$
\r_f^{\P+\E}=S\r_i^{\P+\E}S^\dagger.
\eqno(5)$$
Introduce transition operator $T$ by $S=1+iT$.
{}From unitarity of $S$ we get
\hbox{$T^\dagger T=i(T^\dagger-T)$}; then Eq.~(5) yields [8]:
$$
\Delta\r^{\P+\E}\equiv \r_f^{\P+\E}-\r_i^{\P+\E}
        = {i\over2}[T+T^\dagger,\r_i^{\P+\E}]+
          T\r_i^{\P+\E}T^\dagger
          -{1\over2}\{T^\dagger T,\r_i^{\P+\E}\}.
\eqno(6)$$
The change $\Delta\r$  of the Brownian particle's density operator is given
by the trace of Eq.~(6) over the environmental degrees of freedom.
We shall approximate $d\r/dt$ by
$$
{\Delta\r\over \Delta t}
={1\over\Delta t} tr_{\E}\Delta\r^{\P+\E}
\eqno(7)$$
where $\Delta t$ is the period
spanned by the initial and final states according to Eq.~(5), considered
longer than the typical collision time.

Consider the standard form of the transition
operator for the (spin-independent) scattering of environmental
particle on the Brownian one, with initial laboratory momenta
$\kbi$ and $\pbi$, resp.:
$$
T={1\over2\pi m^\ast}\int d\pbi d\kbi d\kbf f(\kbfa,\kbia)
        \delta(E_{\kbfa}-E_{\kbia})
        |\pbi-\kbfi,\kbf><\pbi,\kbi|.
\eqno(8)$$
This equation is valid in laboratory system though c.m.s. quantities
(marked by stars) appear in it:
$M^\ast=M+m$, $m^\ast=mM/M^\ast$,
$\kbia=(M/M^\ast)\kbi-(m/M^\ast)\pbi$, and
$\kbfa=\kbf-(m/M^\ast)(\kbi+\pbi)$. Also $f$ is the c.m.s. scattering
amplitude.

Assume now the form $\r_i\otimes\re$ for $\r_i^{\P+\E}$. Let the
environment's state $\re$ be stationary,
representing identical particles of
spatial density $n_0$ and of identical momentum distribution $\re(\kb)$.
(Assume Boltzmann statistics, for simplicity's.)
Let us substitute Eq.~(8) into Eq.~(6) and
single out the second term on the R.H.S.:
$$\eqalign{
&tr_{\E}\left(T\r_i\otimes\re T^\dagger\right)
={2\pi n_0\over m^{\ast2}}\int\int d\pbi d\pbi'd\kbi d\kbf \re(\kbi)\cr
&f(\kbfa,\kbia)
 \delta(E_{\kbfa}-E_{\kbia})
  |\pbi-\kbfi><\pbi|\r|\pbi'><\pbi'-\kbfi|
  f^\ast(\kbfap,\kbiap)\delta(E_{\kbfap}-E_{\kbiap}).\cr}
\eqno(9)$$
Change the integration variables $d\kbi,d\kbf$ for $d\kbia,d\kbfa$
respectively.
If $<\pbi|\r|\pbi'>$ is diagonal or {\it almost} diagonal then, to a good
approximation:
$$\eqalign{
tr_{\E}\left(T\r\otimes\re T^\dagger\right)={2\pi n_0\over m^{\ast2}}
        \left(M^\ast\over M\right)^3\int\int &d\pbi d\pbi'd\kbia d\kbfa
                                                   |f(\kbfa,\kbia)|^2
        \left[\delta(E_{\kbfa}-E_{\kbia})\right]^2 \cr
        &\re(\kbi)|\pbi-\kbfi><\pbi|\r|\pbi'><\pbi'-\kbfi|. \cr}
\eqno(10)$$
Rewrite it in operator form:
$$\eqalign{
tr_{\E}\left(T\r\otimes\re T^\dagger\right)=
        &{2\pi n_0\over m^{\ast2}}\left(M^\ast\over M\right)^3\int\int
         d\kbia d\kbfa
         |f(\kbfa,\kbia)|^2
         \left[\delta(E_{\kbfa}-E_{\kbia})\right]^2 \cr
        &\sqrt{\re(\kboi)}e^{-i\kbfia\qbo}\r e^{i\kbfia\qbo}\sqrt{\re(\kboi)}}
\eqno(11)$$
where $\kboi=\kbia+(m/M)(\pbo+\kbfa)$.
Apply the usual approximation
\hbox{$\delta(E)|_{E=0}\sim \Delta t/2\pi$}.
Eq.(11) thus takes the following form:
$${1\over\Delta t}tr_{\E}\left(T\r\otimes\re T^\dagger\right)=
       {n_0\over m^{\ast2}}\left(M^\ast\over M\right)^3
       \int\int d\kbia d\kbfa \delta(E_{\kbfa}-E_{\kbia})
       |f(\kbfa,\kbia)|^2
       V_{\kbfa\kbia}\r V_{\kbfa\kbia}^\dagger
\eqno(12)$$
with
$$
V_{\kbfa\kbia}=
        \sqrt{ \re\left(\kbia+{m\over M}(\pbo+\kbfa)\right) }
        e^{-i\kbfi^\ast\qbo}.
\eqno(13)$$
Now we have completely prepared to calculate the rate
$d\r/dt$ which we approximate by
\hbox{$\Delta\r/\Delta t$} (7).
Consider the R.H.S. of Eq.~(6). The contribution of its first term
will be neglected (that can be justified in rarified gases),
the contribution of the second one is given by
Eq.~(12), and the third term's yield would also be given by a
properly altered form of Eq.~(12). Invoking Eq.~(7), too, all this reads:
$$\eqalign{
{d\r\over dt}={n_0\over m^{\ast2}}\left(M^\ast\over M\right)^3
\int\int &d\kbia d\kbfa \delta(E_{\kbfa}-E_{\kbia})
|f(\kbfa,\kbia)|^2 \cr
         &\left(
          V_{\kbfa\kbia}\r V_{\kbfa\kbia}^\dagger-
          {1\over2}\{V_{\kbfa\kbia}^\dagger V_{\kbfa\kbia},\r\}
          \right).}
\eqno(14)$$
An equivalent equation can be given in terms of the c.m.s. differential
cross section
\hbox{$d\sigma/d\Omega=|f|^2$}:
$$
{d\r\over dt}=n_0\left(M^\ast\over M\right)^3
\int dE^\ast d\Omega_i^\ast d\Omega_f^\ast k^{\ast2}
{d\sigma(\theta^\ast,E^\ast)\over d\Omega_f^\ast}
\left(
V_{\kbfa\kbia}\r V_{\kbfa\kbia}^\dagger-
{1\over2}\{V_{\kbfa\kbia}^\dagger V_{\kbfa\kbia},\r\}
\right).
\eqno(15)$$

This is the central result of our Letter: a quantum master equation
of Lindblad form [3] with Lindblad generators $V_{\kbfa\kbia}$,
describing the fluctuational-frictional evolution
of the Brownian particle's density operator $\r$, valid at time scales
longer than the typical collision time.

The master Eq.~(15) preserves translation invariance:
diagonal (in momentum) density operators remain diagonal.
Eq.(15) implies the following closed equation for translation
invariant states:
$$\eqalign{
{d\r(\pb)\over dt}=&n_0\left(M^\ast\over M\right)^3
\int dE^\ast d\Omega_i^\ast d\Omega_f^\ast k^{\ast2}
{d\sigma(\theta^\ast,E^\ast)\over d\Omega_f^\ast} \cr
&\left(
\re \left({M^\ast\over M}\kbia+{m\over M}(\pbo-\kbfi^\ast)\right)
\r(\pb-\kbfi^\ast)-
\re \left( {M^\ast\over M}\kbia+{m\over M}\pbo \right) \r(\pb)
\right).\cr}
\eqno(16)$$
where, as before, the normalized distribution $\r(\pb)$ is defined
by \hbox{$\r(\pb)=const.<\pb|\rho|\pb>$}. This equation can, at the
same time, be considered the extension of the classical kinetic Eq.~(3)
including friction this time. To find the stationary
solution of the quantum master equation (15) is easy. Since it must be
translationally invariant its diagonal will be subjected to the
classical equation (16).
Invoking the symmetry of $d\sigma$ for $\kbia,\kbfa$ interchanged, the
condition of $d\r/dt=0$ will be the following:
$$
\re (\kbi)\r(\pbi)=\re (\kbf)\r(\pbf).
\eqno(17)$$
i.e. one obtains the {\it detailed balance} condition. It has very
important consequences. Assume the environment is in thermal equilibrium
at temperature $T$, with Boltzmann distribution
$\re(k)=const.\exp(-\beta E)$ where $\beta\equiv1/k_BT$. Then Eq.~(17)
implies that the stationary state $\r$ of the Brownian particle will
be the thermal equilibrium state $const.\exp(-\beta\pb^2/2M)$ in both
classical and quantum cases.

Although various special limits of the general master Eq.~(15)
would be relevant we consider the most traditional
one, i.e., the motion of heavy Brownian particle in
low density Boltzmann gas,
with Ohmic (proportional to velocity) dissipation. In this limiting case
$m/M\rightarrow 0$ but ${mp\over Mk}$ is finite. Then Eq.(15) yields:
$$
{d\r\over dt}=n_0
\int dE d\Omega_i d\Omega_f k^2
{d\sigma(\theta,E)\over d\Omega_f} \re(k_i)
\left(
V_{\kbf\kbi}\r V_{\kbf\kbi}^\dagger-
{1\over2}\{V_{\kbf\kbi}^\dagger V_{\kbf\kbi},\r\}
\right).
\eqno(18)$$

The Lindblad generators (13) will be replaced by:
$$
V_{\kbf\kbi}=
        \left(1-{\beta\over2M}\kbi\pbo\right)
        \exp\left(-i\kbfi\qbo\right).
\eqno(19)$$
The Eq.~(18) with the Lindblad generators (19) belong to the class of
Ohmic dissipative master equations
found phenomenologically by Gallis [9].

Among the early results concerning fenomenological
quantum master equations with friction,
Dekker's proposal [10] is worth to mention.
It is recovered by our Eqs.~(18,19) in the limit
when coherent extension of the Brownian particle and/or transferred
momentum $|\kbfi|$ in single collisions are small enough.
Then $\exp\left(-i\kbfi\qbo\right)$
on the R.H.S. of Eq.~(19) is approximated by $1-i\kbfi\qbo$ and,
neglecting terms of order of $q^2p^2$,
Eq.~(18) can be written as follows:
$$\eqalign{
&{d\r\over dt}=n_0
\int dE d\Omega_i d\Omega_f k^2
{d\sigma(\theta,E)\over d\Omega_f} \re(k_i)
\Bigl(i{\beta\over4M}[\{\kbi\pbo,\kbfi\qbo\},\r]\cr
&+(\kbfi\qbo-i{\beta\over2M}\kbi\pbo)~\r~
      (\kbfi\qbo+i{\beta\over2M}\kbi\pbo)
      -{1\over2}\{(\kbfi\qbo+i{\beta\over2M}\kbi\pbo)
                 (\kbfi\qbo-i{\beta\over2M}\kbi\pbo),\r\}
\Bigr).\cr}
\eqno(20)$$
This is identical to Dekker's fenomenological equation
(in interaction picture):
$$
{d\r\over dt}= -D_{pp}[\qbo,[\qbo,\r]]-D_{qq}[\pbo,[\pbo,\r]]
                 -i{\eta\over2M}[\qbo,\{\pbo,\r\}].
\eqno(21)$$
We can express Dekker's parameters as follows:
$$\eqalignno{
&D_{pp}%={1\over3}<\nu(k_{fi})^2>
       ={4\over3}n_0\int dE d\Omega_i d\Omega_f k^4
       {d\sigma(\theta,E)\over d\Omega_f} \re(k_i)\sin^2{\theta\over2}~~,
                                                                      &(22)\cr
&D_{qq}%={1\over3}\left(\beta\over2M\right)^2<\nu k^2>
       ={1\over3}n_0\left(\beta\over2M\right)^2
        \int dE d\Omega_i d\Omega_f k^4
        {d\sigma(\theta,E)\over d\Omega_f} \re(k_i)~~,&(23)\cr
&\eta  =\beta D_{pp}~.&(24)\cr}
$$
One can now compare the naive quantum master equation (2) to our result
(21-24): we have got an additional term of {\it position diffusion}
definitely absent in the classical equation (1). Classical kinematics, even
if stochastic, is based on continuous trajectories in configurational
space  and does not allow position diffusion at all since this latter
would assume finite position jumps . At the same time, momentum jumps are
reasonable idealizations in classical mechanics. Let us resume
a particular lesson of the longstanding controversies related to the
"true" quantum counterpart of the classical Brownian motion:
quantum friction is accompanied by momentum {\it and position} diffusions,
which latter is
neither present nor even possible in classical kinematics nonetheless
without which no matematically consistent (i.e. Lindblad) quantum
master equation would be written down [11].

This work was supported by OTKA Grant No.1822/1991.
\bigskip\noindent

{}~$^{(a)}$Electronic address: diosi@rmki.kfki.hu

[1] A. O. Caldeira and A. J. Leggett, Physica {\bf 121A}, 587 (1983).

[2] W. H. Zurek, Phys. Today {\bf 44}, No. 10, 36 (1991).

[3] G. Lindblad, Commun. Math. Phys. {\bf 48}, 119 (1976).

[4] V. Ambegaokar, Ber. Bunsenes. Phys. Chem. {\bf 95}, 400 (1991).

[5] W.Pauli, {\it Collected Scientific Papers} (Interscience, New York, 1964).

[6] E. Joos and H. D. Zeh, Z. Phys. {\bf B59}, 223 (1985).

[7] M. R. Gallis, Phys. Rev. {\bf 42A}, 38 (1990).

[8] S. Stenholm, Phys.Scri. {\bf 47}, 724 (1993)

[9] M. R. Gallis, Phys. Rev. {\bf 48A}, 1028 (1993).

[10] H. Dekker, Phys. Rep. {\bf 80}, 1 (1981).

[11] L.Di\'osi, Europhys. Lett. {\bf 22}, 1 (1993); Physica, 199A, 517 (1993).

\vfill\eject
\end